\documentstyle[sprocl,psfig]{article}

\newcommand{\be}{\begin{equation}}
\newcommand{\ee}{\end{equation}}
\newcommand{\bt}{\begin{table}}
\newcommand{\et}{\end{table}}
\newcommand{\btab}{\begin{tabular}}
\newcommand{\etab}{\end{tabular}}

\def\be{\begin{equation}}
\def\ee{\end{equation}}
\begin{document}
\title{PARTICLE PHYSICS, ASTROPHYSICS AND\\[.1cm]
COSMOLOGY WITH FORBIDDEN NEUTRINOS}
\author{R.J. LINDEBAUM, G.B. TUPPER and R.D. VIOLLIER}
\address{Institute of Theoretical Physics and Astrophysics, University of Cape Town,\\
 Rondebosch 7701, Cape, South Africa, e-mail:viollier@physci.uct.ac.za}
\maketitle
\abstracts{Astrophysical and cosmological consequences of a standard 
$\nu_{\tau}$ of (15 $\pm$ 3) keV/$c^{2}$ mass are discussed in the light of the 
recent results of the solar, atmospheric and LSND neutrino experiments and 
theoretical prejudices.}
\section{An exotic neutrino scenario}
Let us assume that the LSND experiment$^{1)}$ is correct and its parameters
$\delta m_{\mu e}^{2} = m_{\nu_{\mu}}^{2} - m_{\nu_{e}}^{2}$
$\approx$ 1 eV$^{2}/c^{4}$ and
$\mbox{sin}^{2}$2$\theta_{\mu e} \approx$ 10$^{-2}$
can be interpreted as $m_{\nu_{\mu}} \approx$ 
1 eV/$c^{2}$ $\gg \; m_{\nu_e}$, in spite of the fact that the KARMEN 
collaboration$^{2)}$ has not observed $\nu_{\mu} \rightarrow \nu_{e}$
oscillations in at least part of the parameter space advocated by LSND. Let us further assume 
that the original quadratic see-saw mechanism based on the up, charm and top 
quark masses$^{3)}$ is the correct explanation for the smallness of the 
neutrino masses, i.e. $m_{\nu} = m_{q}^{2}/M$ with $q = u,c,t$.
Inserting the experimental quark masses  $m_{u} \approx$ 5 MeV/$c^{2}$,
$m_{c} \approx$ 1.5 GeV/$c^{2}$ and $m_{t} \approx$ 180 GeV/$c^{2}$ $\;$ $^{4)}$,
we conclude that the $\nu_{e}$ and $\nu_{\tau}$ masses are  
$m_{\nu_{e}} \approx$ 11.1  $\mu$eV/c$^{2}$ and
$m_{\nu_{\tau}} \approx$ 14.4 $\;$ \mbox{keV}/$c^{2}$,
respectively.
As the $\nu_{\tau}$ mass lies in the cosmologically forbidden region 
between 93 $h^{2}$ eV/$c^{2}$ (0.5 $\leq$  $h$ $\leq$ 0.8)
and 4 GeV/$c^{2}$, we will investigate its cosmological consequences 
below.
Note also that the 
see-saw Majorana mass $M \approx$ 2.25 $\cdot$ 10$^{9}$ GeV
is much smaller than the GUT 
scale $M_{\tiny{GUT}}$ $\approx$ 10$^{16}$ GeV in this scenario.\\[.1cm]
In order to make sure that our choice of neutrino masses does not contradict 
the successful neutrino oscillation interpretation of the solar and 
atmospheric neutrino deficits$^{5)}$ and the $Z^{0}$ width$^{4)}$, 
there must be at least two, but most probably three sterile 
neutrinos $\nu'_{e}, \nu'_{\mu}$ and $\nu'_{\tau}$. Active-sterile 
vacuum oscillations could then account for the solar and atmospheric neutrino 
deficits with $\delta m_{e e'}^{2} = m_{\nu_e}^{2} -
m_{\nu'_e}^{2}$ $\approx 10^{-10}$ eV$^{2}/c^{4}$, $\delta m_{\mu \mu'}^{2} 
= m_{\nu_\mu}^{2} - m_{\nu'_\mu}^{2}$
$\approx$
3 $\cdot$ 10$^{-3}$ eV$^{2}/c^{4}$ and maximal mixing angles for both
$\nu_{e} \rightarrow 
\nu'_{e}$ and $\nu_{\mu} \rightarrow \nu'_{\mu}$ mixing. At this stage it 
is perhaps important to stress the fundamental difference between the maximal 
mixing angles which appear in active-sterile mixing and the small 
flavour mixing sin$^{2}$2$\theta_{\mu e} \approx$ 10$^{-2}$ for 
oscillations between second and first generation.
In fact, if neutrino flavour mixing behaves similar to that of the quark 
sector, we expect third generation mixing angles of about
sin$^{2}$2$\theta_{\tau \mu}$ 
$\approx$ 10$^{-5}$ and sin$^{2}$2$\theta_{\tau e}$ 
$\approx$ 10$^{-9}$, respectively. With such a small mixing angle between third 
and first generation, the $\nu_{\tau}$ is clearly not observable
as a kink in the $\beta$-decay spectrum. Moreover, a Dirac mass of 
$m_{\nu_{\tau}} <$ 30 keV/$c^{2}$ is not in conflict with the duration of
the neutrino burst of SN 1987A$^{25)}$. A $\nu_{\tau}$ mass of 14.4 
keV/$c^{2}$ would, however, 
pose serious problems in neutrinoless $\beta \beta$-decay 
if the $\nu_{\tau}$ was indeed a Majorana neutrino, as required by the
see-saw mechanism. We will thus address this problem below.
Of course, one needs to explain why large angle $\nu_{\mu} \rightarrow 
\nu'_{\mu}$ and $\nu_{\tau} \rightarrow \nu'_{\tau}$ oscillations would 
not spoil the success of Big Bang nucleosynthesis. The effective number of 
neutrinos present during nucleosynthesis could be kept close to 
$N_{eff} \approx$ 3 through matter-enhanced oscillations with large 
neutrino-antineutrino asymmetries just before nucleosynthesis$^{8)}$. 
Alternatively, a phase transition at temperatures around 
$T \approx$ 1 MeV/$k$, associated with the generation of mass differences 
between active and sterile neutrinos, 
could prevent excessive oscillations into the sterile 
sector$^{6)}$.\\[.1cm]
In order to guarantee that the $\nu'_{e}$ and $\nu'_{\mu}$ masses are small and 
nearly degenerate with the $\nu_{e}$ and $\nu_{\mu}$, respectively, 
one requires an extension of the see-saw mechanism to the sterile sector
based the same vacuum expectation value$^{6,7)}$. Apart 
from gravitational interactions and possible neutrino oscillations, sterile 
neutrinos have no interactions in common with particles of the active sector;
they might, however, have interactions with other sterile particles in which 
the particles of the active sector cannot take part. It is not so easy to 
design a consistent model for the sterile particles with interactions
that are renormalisable and anomaly free. In fact, the simplest way to get 
around this problem, is to double the number of particles and assume that 
there is a Standard Model of Particle Physics operating in the sterile sector, as well, as  
e.g. in $E_{8} \times E'_{8}$ superstring theory. The quarks, charged leptons and intermediate bosons of the 
sterile sector would be degenerate with those of the active 
sector, since they are all governed by the same vacuum expectation value
and Yukawa coupling constants which are presumably fixed by a unified theory 
at higher energy. Thus, one of the attractive features of this scenario is
 that parity is 
conserved in both the particle spectrum and the interactions,
with the exception of the neutrino sector. Indeed, since we actually
 observe neutrino oscillations, there must be some small 
breaking of parity symmetry between the active and sterile neutrino 
sectors.\\[.1cm]
Assuming the minimal Higgs sector of one ordinary active Higgs doublet and its 
sterile partner, and restricting ourselves to one generation, the most
general neutrino mass matrix$^{6)}$ can be written in the basis of the 
maximally mixed parity eigenstates $\nu_{L}^{\pm} = 2^{- \frac{1}{2}}
\left( \nu_{L} \pm (N_{R})^{c} \right)$ and $\nu_{R}^{\pm} =
2^{- \frac{1}{2}} \left( \nu_{R} \pm (N_{L})^{c} \right)$ as
\begin{eqnarray}
{\cal{M}} = \left( \begin{array} {cccc} 
0 & 0 & 0 & m_{+}\\[.2cm]
0 & 0 & m_{-} & 0\\[.2cm]
0 & m_{-} & M_{-} & 0\\[.2cm]
m_{+} & 0 & 0 & M_{+} 
\end{array}  \right) \; 
\end{eqnarray}
with $m_{\pm} = m_{1} \pm m_{2}$ and
$M_{\pm} = M_{1} \pm M_{2}$.
In the limit $M_{\pm} \gg m_{\pm}$, the eigenvalues of this mass matrix are 
given by $m_{+}^{2}/M_{+}, m_{-}^{2}/M_{-}, M_{+}$ and $M_{-}$ for the maximally mixed mass and 
parity eigenstates $\nu_{L}^{+}, \nu_{L}^{-}, \nu_{R}^{+}$ and
$\nu_{R}^{-}$, respectively. Specializing to $M_{+} = M_{-} = M$ and
parametrizing $m_{+} = m_{q}$ and $m_{-} = m_{q} - \mu$, the difference of the 
neutrino mass squared is
\begin{equation}
\delta m_{\ell \ell'}^{2} = \frac{m_{q}^{4}}{M^{2}} \left(
1 - \left[ 1 - \frac{\mu}{m_{q}} \right]^{4} \right) \; .
\end{equation}
Assuming that $\mu$ does not depend on the generation because, similar to 
$M$, it is a Majorana mass, we obtain for $\mu$ = 1 MeV/$c^{2}$ reasonable values for 
the parameters of the vacuum oscillation interpretation of the solar and 
atmospheric neutrino deficits and a prediction for $\nu_{\tau}$ oscillations
\begin{equation}
\begin{array} {lcll}
\delta m_{e e'}^{2} &=& 0.73 \cdot 10^{-10} \; \mbox{eV}^{2}/c^{4} \; \; \;
 & \mbox{sin}^{2}2\theta_{e e'} = 1\\[.2cm] \nonumber
\delta m_{\mu \mu'}^{2} &=& 2.7 \cdot 10^{-3} \; \mbox{eV}^{2}/c^{4} \; \; \;
& \mbox{sin}^{2}2\theta_{\mu \mu'} = 1\\[.2cm] \label{3}
\delta m_{\tau \tau'}^{2} &=& 4.6 \cdot 10^{3} \; \mbox{eV}^{2}/c^{4} \; \; \;
 & \mbox{sin}^{2}2\theta_{\tau \tau'} = 1 \; .\\ \nonumber \end{array}
\end{equation}
As $\delta m_{\ell \ell'}^{2}/m_{\nu_\ell}^{2}$ decreases rapidly with increasing neutrino 
mass, the neutrino eigenstates corresponding to low mass eigenvalues may
be combined to form quasi-Dirac spinors, and this would explain why the 
massive $\nu_{\mu}$ and $\nu_{\tau}$ do not make an observable contribution to 
neutrinoless $\beta \beta$-decay.
It is interesting to note that the Majorana mass $\mu$
is approximately equal to the
expected scale of the active-sterile phase transition which we mentioned earlier.
\section{Astrophysical implications}
If a suitable dissipation mechanism exists,
massive neutrinos and antineutrinos will form supermassive compact dark objects in which the 
degeneracy pressure of the neutrinos and antineutrinos balances their 
gravitational attraction. Mass and radius of the most massive object
that can be formed in this 
way are determined by the Oppenheimer-Volkoff (OV) limit$^{9)}$
\begin{eqnarray}
M_{OV} = 0.54195 \left( \frac{\hbar c}{G} \right)^{3/2} m_{\nu}^{-2} \;
g_{\nu}^{-1/2}, \hspace{1cm} R_{OV} = 4.4466 \; R_{OV}^{S}  \; ,
\end{eqnarray}
where $g_{\nu}$ denotes the spin degeneracy factor of the neutrinos and 
antineutrinos 
and $R_{OV}^{S} = 2 GM_{OV}/c^{2}$ is the Schwarzschild radius of 
the mass $M_{OV}$.
There will be equal amounts of right-handed and left-handed neutrinos in the 
neutrino ball, as these are mixed by the gravitational interaction.
There is little difference between a 
supermassive black hole mass and a neutrino ball at the OV-limit, a few 
Schwarzschild radii away from the object, as the radius of the last stable 
orbit around a black hole of the same mass is anyway 3 $R_{OV}^{S}$. Supermassive neutrino 
balls could, therefore, mimic the role of the supermassive black holes which 
are purported to exist at the centres of a large number of galaxies, including 
our own, with masses ranging from 10$^{6.5}$ $M_{\odot}$ to 
10$^{9.5}$ $M_{\odot}$.\\

For instance, if we want to interpret the presumably most massive and violent 
compact dark object in our vicinity, which is located at the centre of\\
M 87 15 Mpc away and has a mass of
 $M$ = (3.2 $\pm$ 0.9) $\cdot$
 10$^{9}$ $M_{\odot}^{12)}$, as a neutrino ball at the OV limit, the 
neutrino mass is constrained by$^{9)}$
\begin{eqnarray}
\begin{array}{lcllcl}
12.4 \; \mbox{keV}/c^{2} &\leq& m_{\nu} &\leq& 16.5 \; \mbox{keV}/c^{2} \; \mbox{for} \; g_{\nu} = 2\\[.2cm]
10.4 \; \mbox{keV}/c^{2} &\leq& m_{\nu} &\leq& 13.9 \; \mbox{keV}/c^{2} \; \mbox{for} \; g_{\nu} = 
4
\end{array}
\end{eqnarray}
which fits our earlier estimate $m_{\nu_\tau} \approx$ 14.4 
keV rather well for $g_{\nu}$ = 2.
A neutrino ball at the OV-limit with a mass $M_{OV}$ = 3 $\cdot$ 10$^{9}$ 
$M_{\odot}$ would have a radius of $R_{OV}$ = 1.52 ld.
Of course, nonrelativistic neutrino balls which are well below the OV limit, 
i.e. $M \ll M_{OV}$, have a size much larger than black holes, 
although they are still dark and much more compact than any known baryonic 
object of the same mass. Mass and radius of nonrelativistic neutrino balls 
scale as$^{19,20,21)}$
\begin{eqnarray}
MR^{3} = \frac{91.869 \; \hbar^{6}}{G^{3} m_{\nu}^{8}} \left(
\frac{2}{g_{\nu}} \right)^{2} \; .
\end{eqnarray}
\noindent
As the gravitational potential of such an extended neutrino ball is much 
shallower, significantly less energy is being dissipated through accreting 
matter than in the case of a black hole of the same mass. In fact, there is a 
compact dark object at the centre of our galaxy which is with $M$ = (2.6 $\pm$ 0.2) 
$\cdot$ 10$^{6}$ $M_{\odot}$ probably at the lower end of the mass spectrum 
for such objects. Its mass is concentrated within a radius smaller than 
0.015 pc or 18 ld$^{13)}$, as determined from the motion of stars in 
the vicinity of the strong radio source SgrA$^{*}$. Interpreting this 
supermassive compact dark object in terms of a neutrino ball,  
the upper limit for the size of the object 
provides us with a lower limit for the neutrino mass, i.e.
$m_{\nu} \geq 15.9 \; \mbox{keV}/c^{2} \; \mbox{for} \; g_{\nu}$ = 2, and
$m_{\nu} \geq 13.4 \; \mbox{keV}/c^{2} \; \mbox{for}\\
g_{\nu}$ = 4, both
corresponding to $M$ = 2.6 $\cdot$ 10$^{6}$ $M_{\odot}$ 
and $R$ = 22.4 ld$^{10,11,22)}$.
In this context, it is important to note that one can, in standard accretion 
disc theory$^{14,23)}$, explain the enigmatic radio and infrared emission 
spectrum  of SgrA$^{*}$ from $\lambda$
$\approx$ 0.3 cm to $\lambda \approx$ 10$^{-3}$ cm 
much better 
in the neutrino ball scenario than in the black hole scenario.
For a fit of the spectrum with a neutrino ball, the neutrino mass should be 
between 12 keV/$c^{2}$ and 18 keV/$c^{2} \;^{14,23)}$ for $g_{\nu}$ = 2.
We thus conclude that there are compelling astrophysical arguments for the 
existence of a neutral weakly interacting fermion of mass $m_{\nu}$ = (15 $\pm$
3) keV/$c^{2}$ which could be the $\nu_{\tau}$.
\section{Cosmological implications}
\begin{figure}[!t]
\psfig{figure=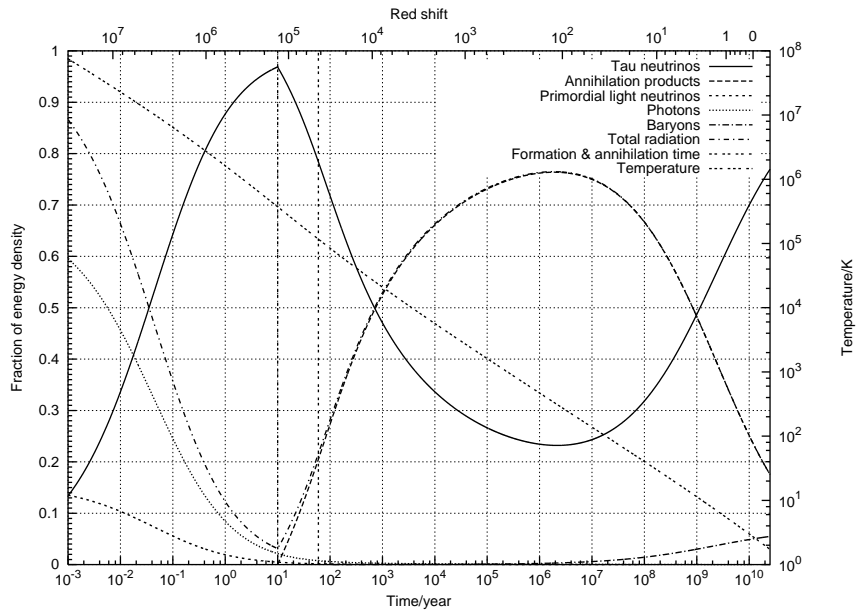} \label{energyden}
\caption{Contributions of the various particle species to the critical 
density as a function of time, for a Hubble parameter $h$ = 0.5, age of the
universe $t_{0}$ = 11.7 Gyr, formation time
$t_{f}$ = 10 yr and annihilation time scale $t_{a}$ = 50 yr.} 
\end{figure} 
To be specific, we now assume that the $\nu_{\tau}$ is a standard Dirac 
neutrino with mass $m_{\nu_\tau}$ = 14.4 keV/$c^{2}$. Such a neutrino is 
quasi-stable, as its lifetime is many orders of magnitude larger than the age 
of the universe$^{21)}$. The $\nu_{\tau}$ will be non-relativistic 54 min 
after the Big Bang at a photon temperature $T_{NR}^{\gamma}$ = 20 keV/$k$,
long after nucleosynthesis. The universe will become heavy neutrino matter
dominated 22 d after the Big 
Bang at $T_{E}^{\gamma}$ = 1 keV/$k$. From now on the evolution of the 
universe will differ substantially from that of the Standard 
Model of Cosmology$^{24)}$. In fact, based on the Thomas-Fermi model at finite 
temperature, it has been shown$^{15,16,17,18)}$ that some time during the heavy 
neutrino matter dominated epoch, the universe 
will undergo a {\it gravitational} phase transition, yielding a condensed phase 
that consists of quasi-degenerate supermassive neutrino balls with masses 
close to the OV-limit $M_{OV}$ = 3 $\cdot$ 10$^{9}$ $M_{\odot}$.
Of course, for this phase transition to happen,
one would need an efficient dissipation 
mechanism which could be based on nonstandard bosons associated with 
the 1 MeV phase transition mentioned earlier. 
The time at 
which the first-order gravitational phase transition starts, depends on the detailed model, 
but it will typically begin 1 to 2 yr after the Big Bang.
The first neutrino balls will be formed about 10 yr after Big Bang, which allows the neutrinos to 
condense
in the 
neutrino balls in the free-fall time. Only a fraction of
10$^{-3}$ of the neutrinos are estimated to be in the gaseous phase
after this 
phase transition leading to a neutrino dominated critical universe today.
The latent heat released is about 3.6\% of the rest mass of the neutrino 
balls$^{9)}$.
Soon after the formation of the neutrino balls, annihilation of the heavy 
neutrinos into nonstandard light
bosons will take place 
efficiently in the dense interior of the neutrino balls$^{21)}$
reducing the neutrino number density.
Due to e.g. S-wave annihilation the mass 
of the neutrino balls will decrease as $^{21)}$
$M(t) = M_{0}\left[ t_{a}/(t_{a} + t - t_{f}) \right]^{\frac{1}{2}}$, where
the annihilation time $t_{a}$ is determined by the annihilation rate
and $t_{f}$ is the formation time of the neutrino balls. 
%%FIGURE, etc.
In Fig.1 we have plotted for a critical universe the fraction of the total energy 
density that the various particles make up as a function of time.
The annihilation products will start to dominate the energy density of the universe 
around 650 yr. Recombination will take place in this radiation dominated 
phase around 31 kyr or $z \approx$ 1100.  Once the $\nu_{\tau}$ or $\bar{\nu}_{\tau}$
in the neutrino balls have annihilated below a certain level,
and the energy density of the annihilation products has cooled 
sufficiently, the dispersed $\nu_{\tau}$ and $\bar{\nu}_{\tau}$  
that escaped the phase transition, will start dominating the universe
around 1 Gyr again, thereby igniting the quasars through accretion.
Of course, annihilation of the
$\nu_{\tau}$ and  
$\bar{\nu}_{\tau}$ in the neutrino balls will stop as soon as either the
$\bar{\nu}_{\tau}$ or the $\nu_{\tau}$ are depleted. A $\nu_{\tau}$ - $\bar{\nu}_{\tau}$
asymmetry of about 10$^{-3}$ would be consistent with masses of the
neutrino balls today. 
In summary, it is refreshing to see that the desert of the Standard Model of 
Cosmology between nucleosynthesis and recombination is being revived in this 
scenario. 

\end{document}